\documentclass[11pt,twoside]{article}
\usepackage[hmarginratio=1:1,top=32mm,columnsep=20pt]{geometry}
\usepackage{multicol}
\usepackage[footnotesize,labelfont=bf,up,textfont=it,up]{caption}
\usepackage{float}
\usepackage{booktabs}
\usepackage{paralist}
\usepackage{amsmath}
\usepackage{amsthm}
\usepackage{wasysym}
\usepackage{graphicx}
\usepackage{hyperref}

\usepackage{abstract}

\usepackage{titlesec}
\renewcommand\thesection{\Roman{section}}
\renewcommand\thesubsection{\Roman{subsection}}
\titleformat{\section}[block]{\large\scshape\centering}{\thesection.}{1em}{}
\titleformat{\subsection}[block]{\large}{\thesubsection.}{1em}{}

\usepackage{fancyhdr}
\title{\textsc{Measuring Prestige in Online Social Networks}}
\author{Stan Palasek\\Department of Mathematics, Princeton University}
\date{July 25, 2014}

\newtheorem{theorem}{Theorem}
\newtheorem{definition}[theorem]{Definition}

\begin{document}

\maketitle

\thispagestyle{fancy}

\begin{abstract}
We study the locally-defined social capital metric of Palasek (2013) for determining individuals' prestige within an online social network. From it we derive an equivalent global measure by considering random walks over the network itself. This result inspires a novel expression quantifying the strategic desirability of a potential social connection. We show \textit{in silico} that ideal social neighbors tend to satisfy a ``big fish in a small pond'' criterion and that the distribution of neighbor-desirability throughout a network is governed by anti-homophily.
\end{abstract}

\section{Introduction}

Economic analysis of digital commodities is often made difficult by the absence of any marginal cost to the distributor~\cite{bakos}. Although some goods maintain their propriety, many others--open-source software being a widely-studied example--are distributed online at no direct cost to the consumer. There are competing theories to explain the latter seemingly-uneconomic behavior including, for instance, an expectation of reciprocity~\cite{veale} and reputation-seeking by the producers~\cite[p.~53]{raymond}. In either case, in order to begin to study the producer's decision-making process, we must construct a theory of utility which justifies potentially significant fixed costs, the details of which will likely depend on the peculiarities of the market at hand.

In this work we will focus our attention on the exchange of social gratification in online communities. This situation evidently falls into the category described above in which a commodity is exchanged but at no marginal cost to the supplier. The literature has indeed identified the acquisition of social capital as a central motivation for social media use~\cite{ellison}. We will proceed using the utility function introduced in \cite{palasek}, to be recalled in the next section. This formulation offers a local method of measuring the centrality of a member of a social network based on the rate at which he or she receives ``likes,'' calculated relative to the centralities of the individual's social neighbors. In the author's previous work, the equilibrium effects of prestige-seeking across the social network were considered. Here we will work with the mathematical formalism itself to gain deeper insight into how it is calculated (for its definition is recursive) and to identify related metrics which may be sociologically relevant.

\section{Defining a ``like''-based centrality measure}

We will consider an online social network over an undirected graph $G$ of order $N$. Agents may bestow ``likes'' upon each other with a non-negative rate. Labeling $G$'s vertices from $1$ to $N$, we let $R$ be the $N\times N$ matrix in which the entry $R_{i,j}$ represents the rate at which $j$ ``likes'' $i$. We require that no transactions occur between a pair for which there is no edge in $G$; ie.\ if $(i,j)\notin G$ then $R_{i,j}=0$.

Next we will define a centrality measure similar in concept to the random walk betweenness centrality of Newman \cite{newman}.
\begin{definition}\label{p}
Let $G$ be a connected undirected graph of order $N$ with nodes of degree $\delta_1,\delta_2,\ldots,\delta_N$. Define a matrix
\begin{align}
P=\textnormal{diag}(\delta_1^{-1},\delta_2^{-1},\ldots,\delta_N^{-1})G
\end{align}
where $\textnormal{diag}(\cdot)$ is the diagonal matrix with the given entries and $G$ here is the adjacency matrix of the graph $G$.
We will denote as $\mathbf{p}$ the normalized eigenvector of $P$ corresponding to the eigenvalue of maximum magnitude as given by the Perron-Frobenius theorem.
\end{definition}
It is easy to verify that $P$ is a stochastic matrix. Repeated multiplication by it therefore models the Markov chain of a memoryless random walk over the graph $G$. The eigenvector $\mathbf{p}$, the steady-state probability distribution over the vertices of $G$, will become of use to us later. We will now define another centrality measure, this time using the matrix $R$ as well. We proceed in analogy to the well-known eigenvector centrality, but now scaling by the centralities of the neighboring nodes. Such an adjustment simulates the effect of an individual's centrality appearing greater or lesser when compared to those with whom he interacts.
\begin{definition}[likedness centrality from \cite{palasek}]\label{lcdef}
Let $G$ be an undirected graph of order $N$ and $R$ a real $N\times N$ matrix with the criteria above. Furthermore, let a likedness centrality vector $\mathbf{L}$ be any vector of positive components satisfying the recursive relation
\begin{align}\label{lc}
\sum_{j=1}^N(G_{i,j}L_i-R_{i,j})L_j=0
\end{align}
where $G_{i,j}$ is the $(i,j)$ entry of $G$'s adjacency matrix.
\end{definition}
We will often identify the coordinates of this vector with the respective vertices of $G$. Note that \eqref{lc} is unaffected up to a scaling of $\mathbf{L}$ and $R$ so this definition cannot be unique. We remedy this as follows.
\begin{definition}[unique likedness centrality]
Define the \textit{unique} likedness centrality to be the vector $\mathbf{L}^*$ satisfying both \eqref{lc} and
\begin{align}
\mathbf{p}\cdot\mathbf{L}^*=1
\end{align}
where $\mathbf{p}$ is as in Definition \ref{p}.
\end{definition}
If $G$ is connected as specified in Definition \ref{p}, then the components of $\mathbf{p}$ are nonzero and uniqueness can be achieved simply by normalization.

\section{A random walk interpretation of $\mathbf{L^*}$}

It will greatly simplify our notation henceforth to define for a non-isolated node $i$
\begin{align}
\sigma_i=\sum_{j=1}^NG_{i,j}L_j.
\end{align}
Thus we can rearrange \eqref{lc} to obtain
\begin{align}
L_i=\frac{1}{\sigma_i}\sum_{j=1}^NR_{i,j}L_j,
\end{align}
recalling that Definition \ref{lcdef} disallows $L_j$, and therefore $\sigma_i$, to be zero. Iteratively applying this formula $n$ times yields
\begin{align}
L_{i_0}=\sum_{i_1=1}^N\sum_{i_2=1}^N\cdots\sum_{i_n=1}^N\left(L_{i_n}\prod_{j=0}^{n-1}\sigma_{i_j}^{-1} R_{i_j,i_{j+1}}\right).
\end{align}
The product is only nonzero when all of the $R_{i_j,i_{j+1}}$ are positive, which we specified occurs only when all of the edges $(i_j,i_{j+1})$ are included in $G$. Therefore it is sufficient to sum over all paths on $G$ of length $n$ beginning at $i_0$. That is,
\begin{align}
L_{i_0}=\textsc{(number of $n$-paths on $G$ from $i_0$)}\times\left\langle L_{i_n}\prod_{j=0}^{n-1}\sigma_{i_j}^{-1}R_{i_j,i_{j+1}}\right\rangle
\end{align}
where $\langle\,\cdot\,\rangle$ is the mean over all $n$-paths on $G$ of the form $(i_0,i_1,\ldots,i_n)$. For now we will refer to the number of $n$-paths from $i_0$ as $\pi_{i_0}^n$. To work with the second factor, we will consider what happens when $n$ becomes large. We begin by rearranging the product into the form
\begin{align}
L_{i_0}=\lim_{n\to\infty}\pi_{i_0}^n\left\langle L_{i_n}\prod_{i=1}^N\left(\sigma_i^{-\#i}\prod_{j:(i,j)\in G}R_{i,j}^{\#(i,j)}\right)\right\rangle
\end{align}
where $\#i$ and $\#(i,j)$ are respectively the number of times the vertex $i$ and the edge $(i,j)$ appear in the path. We call upon the law of large numbers to reinterpret the expectation operator as a mean over many random paths beginning at $i_0$. The product over $i$ weighs all steps of the walk equally so for $n\gg1$, the choice of $i_n$ and therefore the value of $L_{i_n}$ approaches independence from the product. This independence permits us to invoke multiplicativity of the expectation operator. We defined the transition matrix for a Markov walk over a graph in Definition \ref{p} whence we can evaluate $\langle L_{i_n}\rangle$ exactly.
\begin{align}
L_{i_0}=\lim_{n\to\infty}\pi_{i_0}^n\left(\sum_{i=1}^N(\mathbf{i_0}\!\!^\top P^n)_iL_i\right)\left\langle\prod_{i=1}^N\left(\sigma_i^{-\#i}\prod_{j:(i,j)\in G}R_{i,j}^{\#(i,j)}\right)\right\rangle
\end{align}
Furthermore, the distribution of vertices follows the probability distribution resulting from the Markov random walk on $G$ for which $P$ is the transition matrix. Therefore, noting that the frequency of an edge is the frequency divided by the degree of the starting vertex, we have
\begin{align}
L_{i_0}=&\lim_{n\to\infty}\pi_{i_0}^n(P^n\mathbf{i_0})\cdot\mathbf{L}\prod_{i=1}^N\left(\sigma_i^{-\left(\mathbf{i_0}\!\!^\top\sum_{t=0}^{n-1}P^t\right)_i}\prod_{j:(i,j)\in G}R_{i,j}^{\left(\mathbf{i_0}\!\!^\top\sum_{t=0}^{n-1}P^t\right)_i/\delta_i}\right).
\end{align}
Observe that we can also express the number of $n$-paths in terms of a sum of powers of $P$. In particular, the location at each step is a probability distribution over the graph's nodes derived from the same Markov process as before. The graph's degree distribution then gives the number of possible transitions from each vertex. Thus,
\begin{align}
\pi_{i_0}^n=\prod_{t=0}^{n-1}\prod_{i=1}^N\delta_i^{(\mathbf{i_0}\!\!^\top P^t)_i}=\prod_{i=1}^N\delta_i^{\left(\mathbf{i_0}\!\!^\top\sum_{t=1}^{n-1}P^t\right)_i}
\end{align}
which we can rearrange to fit into our expression for $L_{i_0}$.
\begin{align}
L_{i_0}=(\mathbf{p}\cdot\mathbf{L})\lim_{n\to\infty}\prod_{i=1}^N\left[\left(\frac{\sigma_i}{\delta_i}\right)^{-1}\cdot\left(\prod_{j:(i,j)\in G}R_{i,j}\right)^{1/\delta_i}\right]^{\left(\sum_{t=0}^n P^t\right)_{i_0,i}}
\end{align}
This result is far more intuitive than it may appear, although the manner in which it is arranged may hint at some simplifications. The scaling factor is that which we required to define the unique likedness centrality. Furthermore, recalling the definition of $\sigma_i$, we can interpret the first term in the product, $\sigma_i/\delta_i$, to be the mean likedness of the nodes connected to $i$. Finally, the second term in the product is simply the geometric mean of the rates at which $i$ is ``liked'' by its connections. Thus we have, adjusting the indexing,
\begin{align}\label{eq}
L_i^*=\lim_{n\to\infty}\prod_{j\in V(G)}\left(\frac{\text{\textsc{GM of rates at which $j$ is ``liked''}}}{\text{\textsc{AM of $j$'s connections' likednesses}}}\right)^{\left(\sum_{t=0}^n P^t\right)_{i,j}},
\end{align}
recalling our earlier assumptions which provide that $\mathbf{p}\cdot\mathbf{L}\neq0$. The astute reader might notice that because $P$ is stochastic, the Perron-Frobenius theorem guarantees that it has an eigenvalue of 1 and its geometric series does not therefore converge. Although this is true, some experimentation reveals that despite the exponents' being unbounded, the bases' distribution both above and below unity leads to convergence of the product itself.

\section{Measuring neighbors' desirability}

The result of the previous section provides a qualitatively different interpretation as to how likedness centrality is calculated. We originally defined it as a local measure of the prominence of a node relative to his immediate neighbors. Instead, we now define a score for every vertex in the network based on its local prominence and incorporate each, scaled by its closeness to $i$, into the computation $L_i^*$. This score is in a convenient form in \eqref{eq} as it appears independently of both the length of the walk and the agent whose centrality is being calculated. It is therefore sensible to study it as a property of a node in its own right.
\begin{definition}[neighbor desirability]
Define the neighbor desirability of a non-isolated vertex $i$ to be the ratio of the geometric mean of the rates at which $i$ receives ``likes'' to the arithmetic mean of the likedness centralities of those with whom $i$ is connected.
\end{definition}
The equations derived in the last section illustrate that this ratio quantifies the desirability of being near $i$ in the network. Neighbor desirability resembles likedness centrality in all ways but the fact that the former metric decreases strictly with respect to the likedness centrality of the neighboring nodes, whereas the effect is non-strict for the latter. More explicitly, it is most desirable to be near those who, though highly ``liked,'' are among those with little prestige themselves. This ``big fish in a small pond'' observation is provided robustly by \eqref{eq} despite the complexity resulting from heavy interdependence of the centralities throughout a community. The emergence of a geometric, not arithmetic, mean of rates may come as a surprise. This is not merely a formal distinction; the geometric mean uses a far more egalitarian weighing of the samples. That is, even if a node $i$ ``likes'' at a far greater rate than another $j$ does, perturbing either rate by a fixed proportion of its starting value has an equivalent effect on the result of the mean.

\begin{figure}
\begin{centering}
\includegraphics[scale=.75]{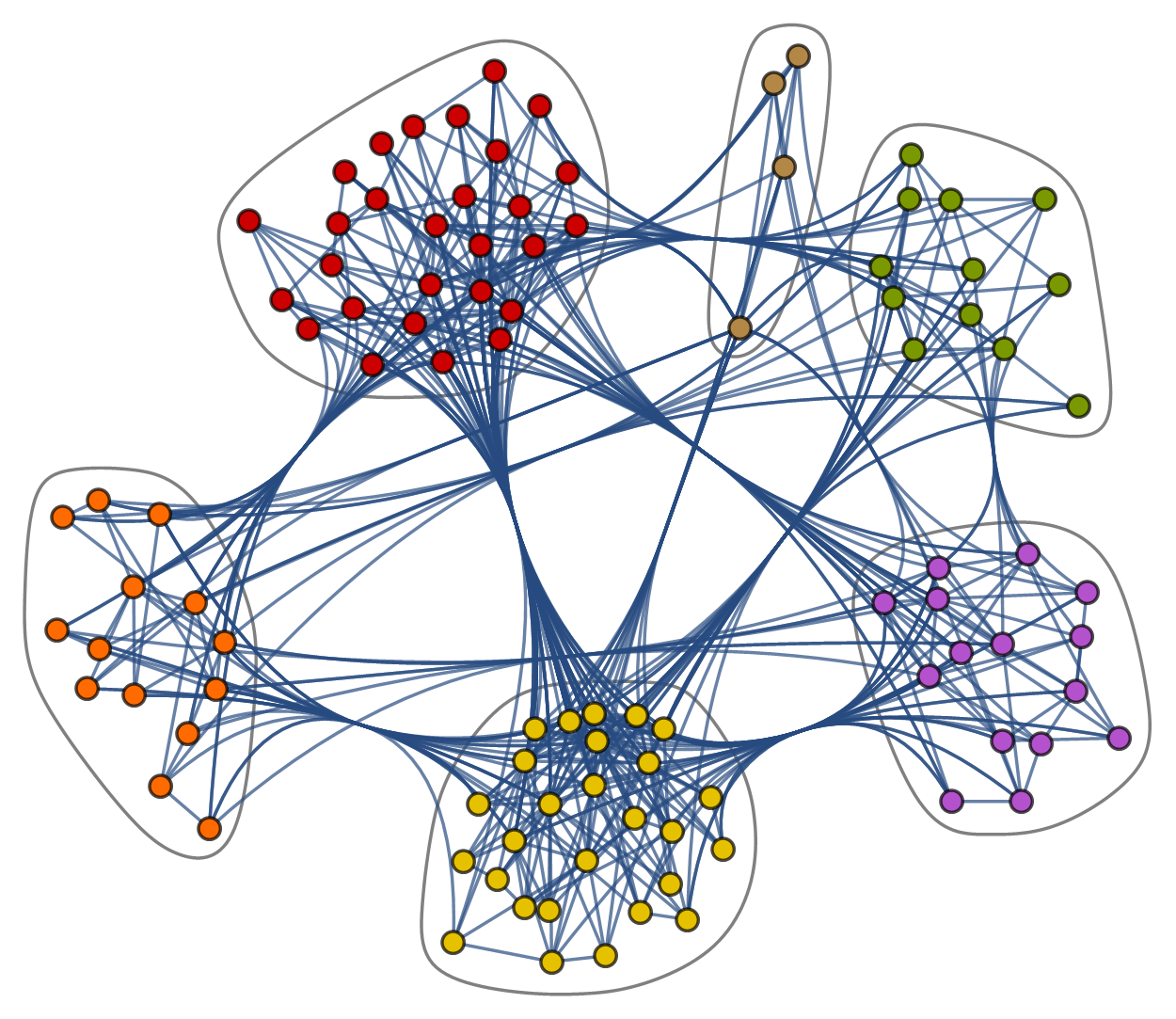}
\caption{The sample graph $G$, generated with the Barabasi-Albert model with parameters $m_0=5$ and $t=95$ (employing the authors' notation). Communities are identified according to a modularity-based clustering metric.}\label{g}
\end{centering}
\end{figure}

In order to examine the structural factors of a network that may affect a vertex's neighbor desirability, we conduct a computer experiment as follows. We randomly select a graph $G$ of order 100 according to the Barabasi-Albert random graph distribution, the result of which is shown in Figure \ref{g}. This model is known to produce networks with realistic characteristics including preferential attachment (yielding a scale-free degree distribution) and low average path length~\cite{barabasi}. For this fixed graph we randomly generate 10,000 ``like'' rate ensembles and compute the likedness centrality vector in each case, effectively yielding 1,000,000 distinctly operating individuals to study. The rates of ``liking'' are assumed to follow an exponential distribution with $\lambda=1/2$.

\begin{figure}
\includegraphics[scale=.22]{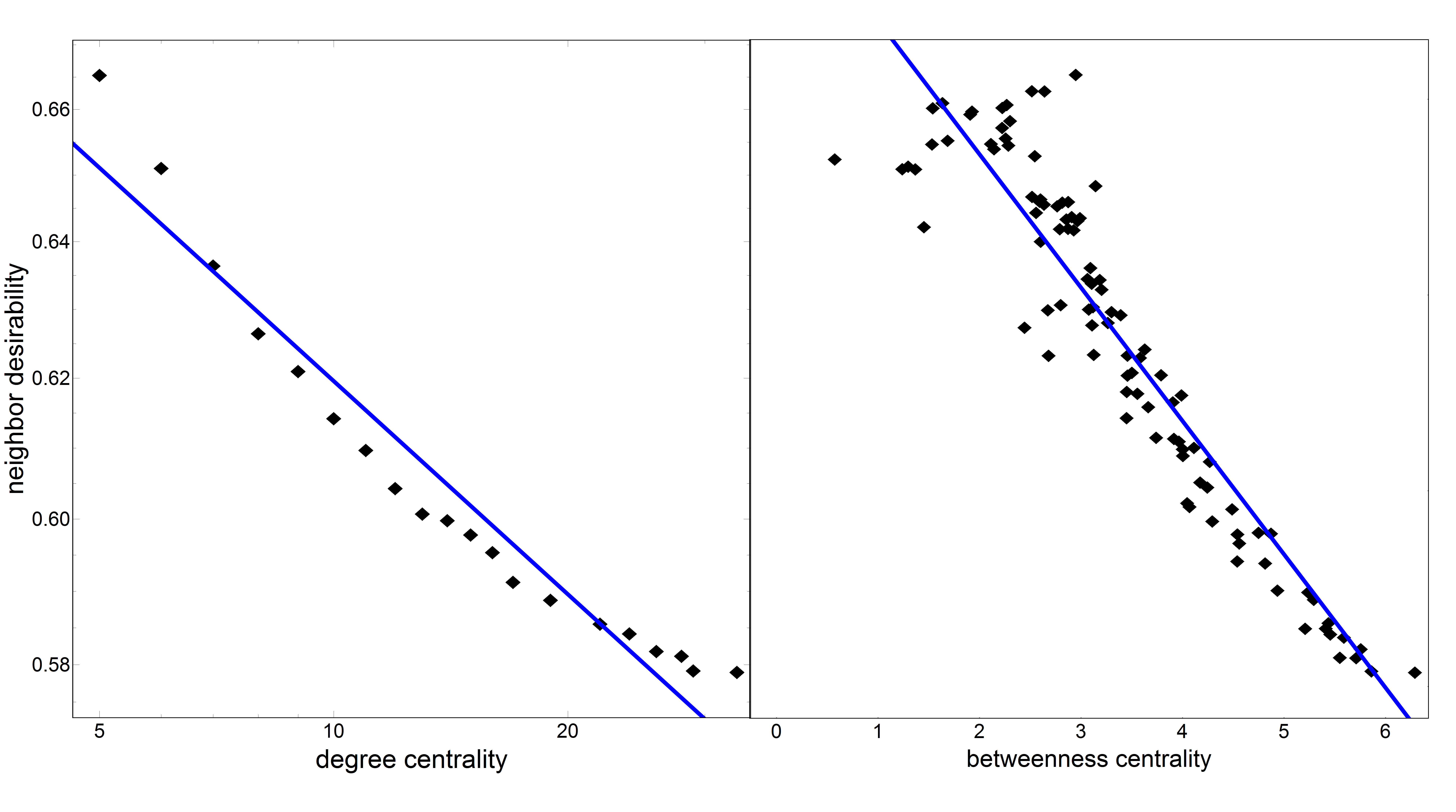}
\caption{Neighbor desirability's dependence on some traditional centrality measures, computed over 10,000 samples of communities populating the graph $G$. Robust negative trends are identified on the log-log scale. In particular, the relationship between a node's neighbor desirability and betweenness centrality is closely given by $\text{N\!D}=0.708\times\text{BC}^{-0.034}$ ($R^2>0.999$).}\label{centralities}
\end{figure}

It should be noted that although \eqref{eq} appears straightforward, the denominator in the product depends on a number of structural features of the graph as demonstrated in \cite{palasek}. We therefore examine the effect of various centrality measures on the neighbor desirability. Because all the standard metrics prove to exhibit a qualitatively similar trend, only those with the least noise are shown in Figure \ref{centralities}. Consistent with the latter part of our prior ``big fish in a small pond'' criterion for neighbor desirability, well-connected nodes suffer a clear deficiency in desirability, with depletion most nearly resembling a power law.

\begin{figure}
\begin{centering}
\includegraphics[scale=.22]{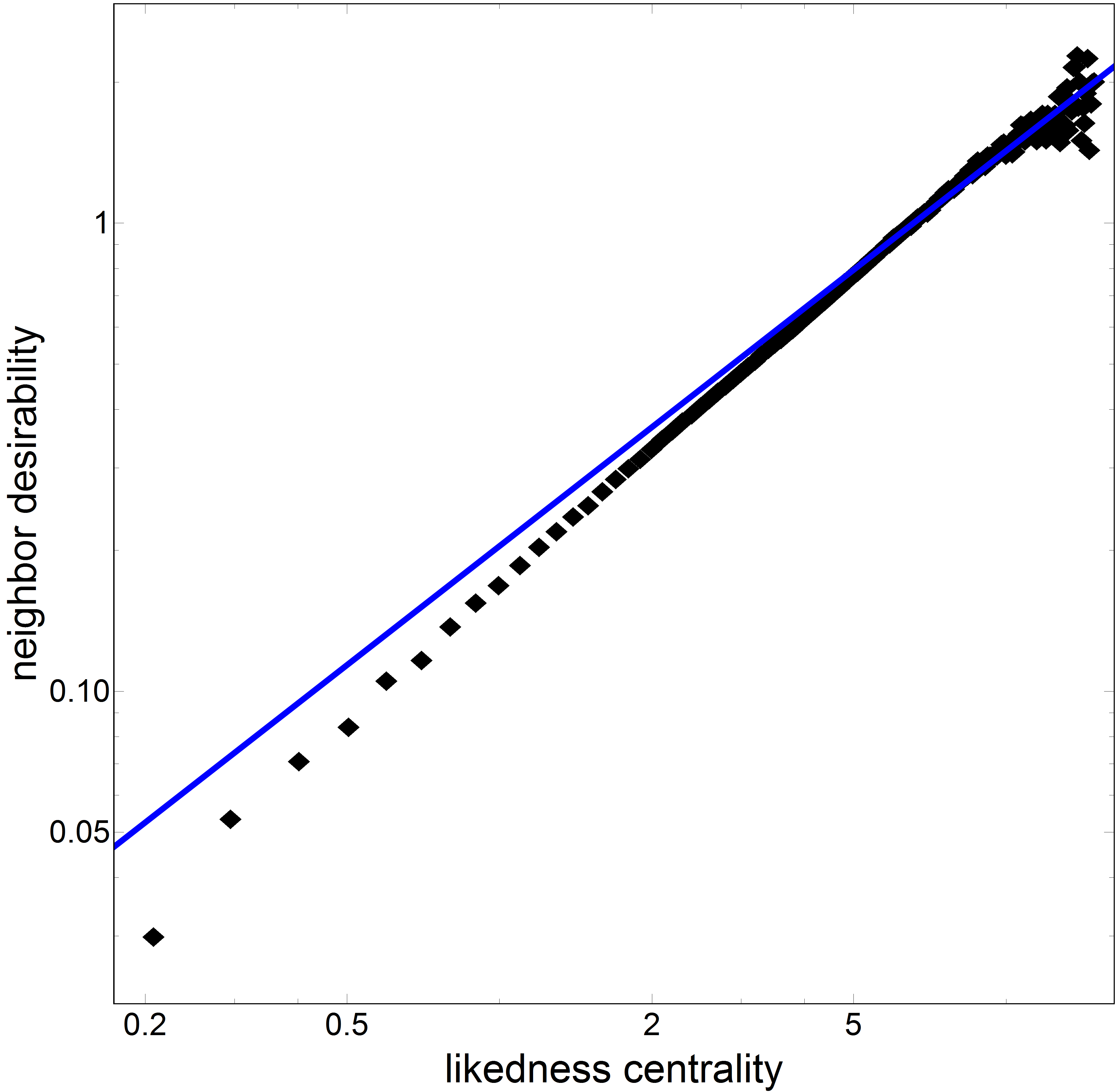}
\caption{Neighbor desirability's dependence on mean likedness centrality. The log-log scale reveals an asymptotic power law trend, inflicted by some heteroscedasticity due to the tail of the population-wide distribution of likedness centrality. The precise relationship is given by $\text{NC}=0.204\times\text{LC}^{0.844}$ ($R^2>0.99$).}\label{corr}
\end{centering}
\end{figure}

To investigate the ``big fish'' effect we might also study the relationship between an agent's likedness centrality and neighbor desirability. This may seem like a trivial consideration given that desirability is proportional to the mean of the incoming rates, but that is to neglect the reciprocal nature of likedness centrality. Nonetheless, a strong positive correlation between the two measures of centrality is found and depicted in Figure \ref{corr}.

\begin{figure}
\begin{centering}
\includegraphics[scale=.22]{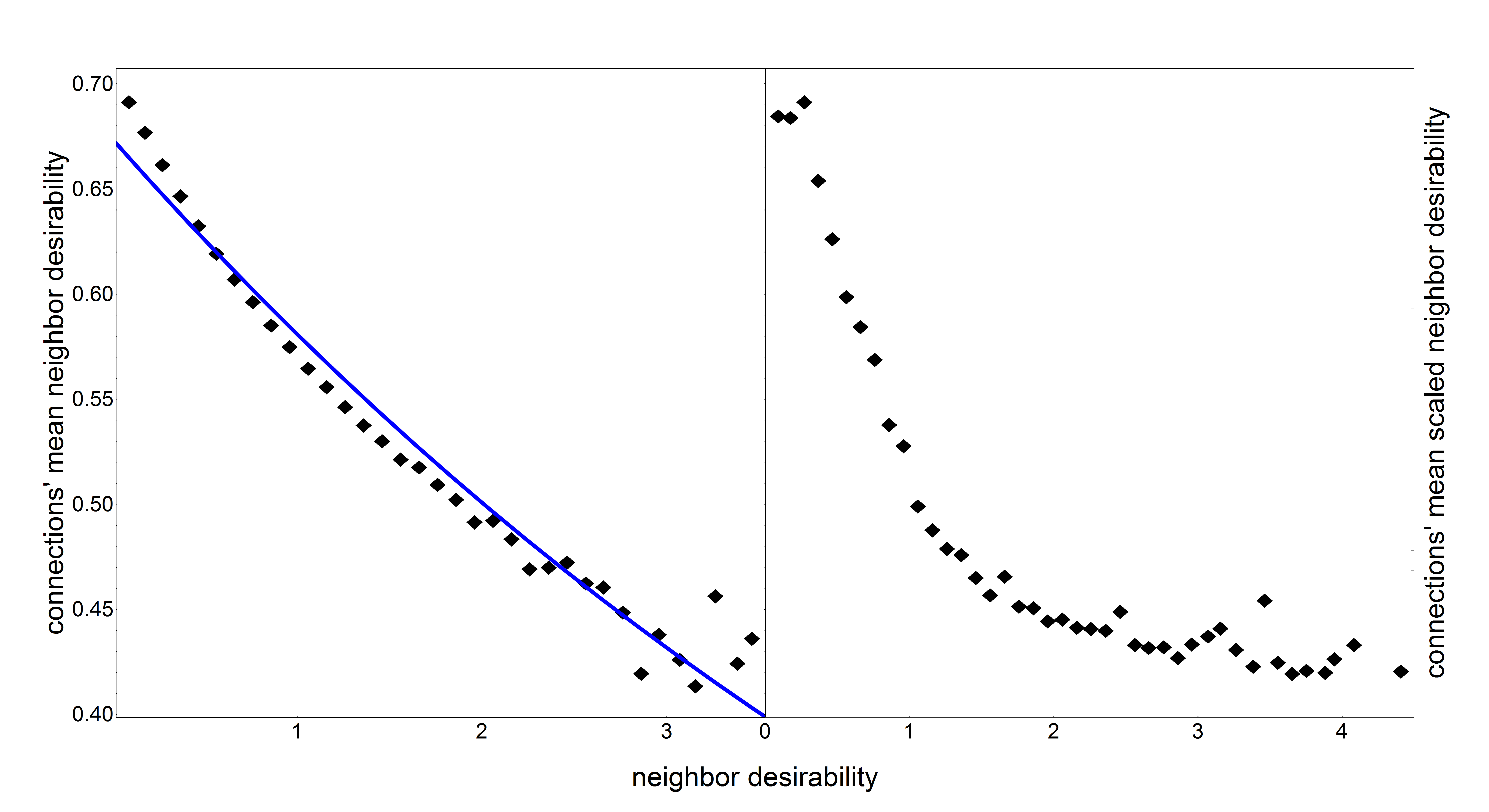}
\caption{The effect of neighbor desirability on the mean desirability of one's connections. The two log-plots respectively show this correlation in absolute terms and scaled to show how over-represented a neighbor desirability is with respect to the global distribution. The unweighted version is plotted along with the exponential fit $\langle N\!D\rangle=0.674\times\exp(-0.148N\!D)$ ($R^2>0.999$).}\label{pairs}
\end{centering}
\end{figure}

\begin{figure}
\begin{centering}
\includegraphics[scale=.75]{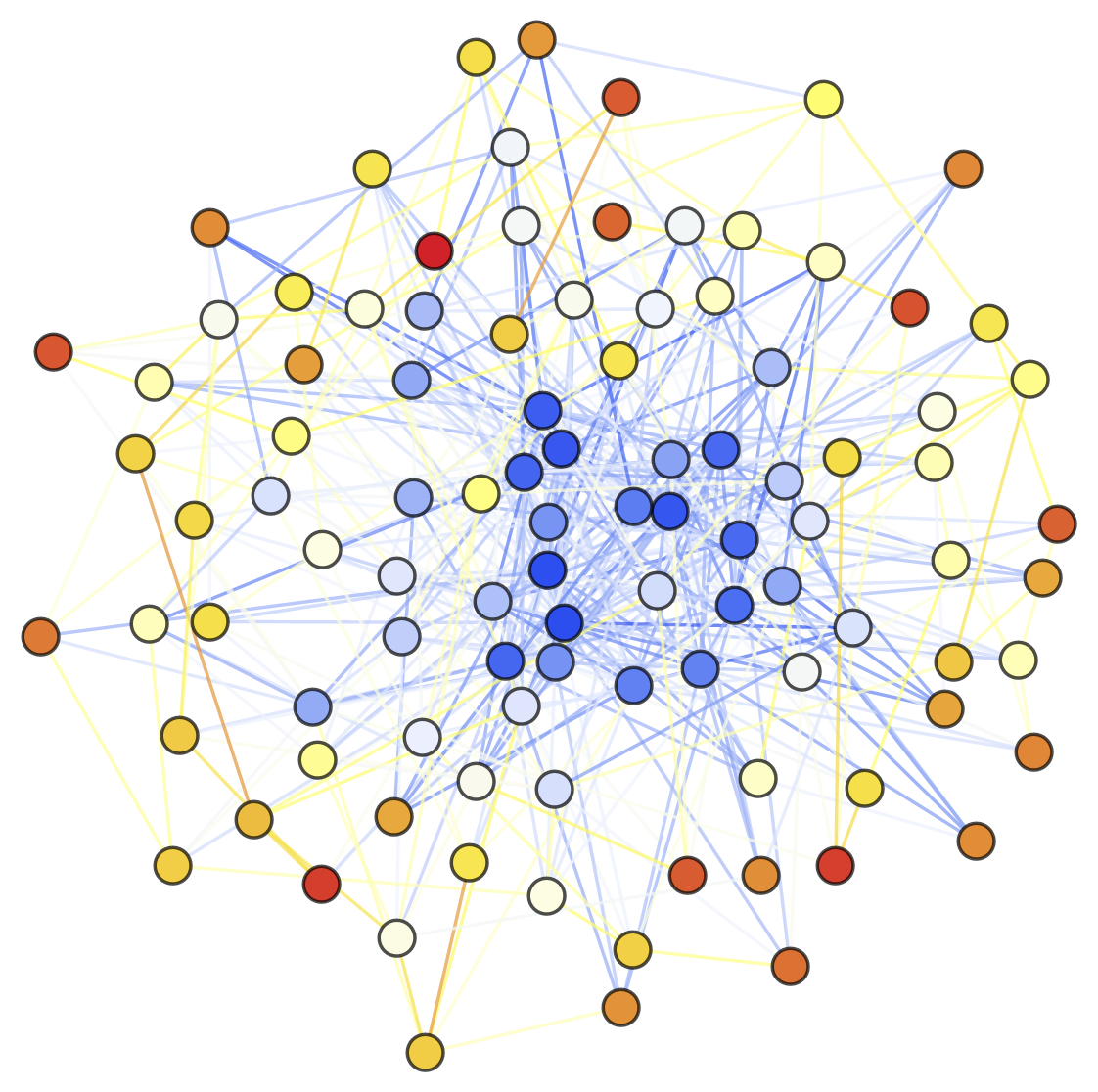}
\caption{The graph $G$ is again plotted, now with vertices colored by their neighbor desirability (increasing from blue to red). The previously-observed effects including inverse correlation with centrality and anti-homophily are apparent. A neutral node (of unitary neighbor desirability) appears as yellow ($\text{RGB}\approx(1,1,.4)$).}\label{colors}
\end{centering}
\end{figure}

Finally, we wish to determine whether connected pairs of nodes are more or less likely to have like neighbor desirabilities. The first panel in Figure \ref{pairs} illustrates that the most desirable neighbors are themselves proximate to worse neighbors on average. This trend becomes even more pronounced when we take into account the sampling bias introduced by the underlying distribution of neighbor desirability, shown in the second panel of the same figure. This observation and the prior regarding centrality measures become apparent in Figure \ref{colors} which shows $G$ with its vertices and edges colored according to their neighbor desirability and mean neighbor desirability, respectively.

\section{Conclusion}

Our first key finding is \eqref{eq}, an equivalent formulation of Definition \ref{lc}. This result is remarkable in that it begins with likedness centrality in a quasi-eigenvector centrality form, which only directly depends on a node's neighbors, and produces a global metric, which considers a characteristic of every vertex in the graph and scales them according to their proximities to the vertex at hand. This characteristic is found to be useful in and of itself as a measure of a node's effect on the social prestiges of others. We were able to identify several robust properties of this so-called ``neighbor desirability.'' First, distinct from likedness centrality, it correlates negatively with traditional structural centrality metrics (Figure \ref{centralities}). Then, as would be expected, we saw that desirable neighbors have more social capital on average (Figure \ref{corr}). Finally, desirable nodes tend to have connections with less desirable nodes on average (Figure \ref{pairs}). These three factors combine to what we call the ``big fish in a small pond'' criterion according to which the most desirable neighbors tend to be very prominent but within less-esteemed circles.

We conclude with a remark on the limitations of this study's methods. Throughout the \textit{in silico} experiments of Section IV, we assumed that the entries of $R$ are independent and identically-distributed, neglecting any strategic behavior by the agents. Therefore, we may only consider this to be a study of the neighbor desirability metric itself and not of features of realistic communities which respond to it and its derivatives. Strategic optimization of likedness centrality in particular has been investigated in \cite{palasek}, which shows that societies of prestige-seeking agents can produce some emergent network features. A similar analysis must be done for neighbor desirability, though the problem of finding equilibrium rate matrices is made difficult by the large number of variables and nonlinearity of likedness centrality.

\bibliographystyle{unsrt}

\begin{thebibliography}{9}

\bibitem{bakos}
Bakos, Yannis.\ (1998). The emerging role of electronic marketplaces on the Internet. \textit{Communications of the ACM, 41}(8), 35-42.

\bibitem{veale}
Veale, K.\ (2003). Internet gift economies: Voluntary payment schemes as tangible reciprocity. \textit{First Monday, 8}(12).

\bibitem{raymond}
Raymond, E.\ S.\ (2001). \textit{The Cathedral \& the Bazaar: Musings on Linux and Open Source by an Accidental Revolutionary.} O'Reilly Media, Inc.

\bibitem{ellison}
Ellison, N.\ B., Steinfield, C.,\ \& Lampe, C.\ (2007). The Benefits of Facebook ``Friends:'' Social Capital and College Students' Use of Online Social Network Sites. \textit{Journal of Computer-Mediated Communication, 12}(4), 1143-1168.

\bibitem{palasek}
Palasek, S.\ (2013). On the Strategic Allocation of Social Gratification. \textit{arXiv:1309.2052.}

\bibitem{newman}
Newman, M. E.\ (2004). A measure of betweenness centrality based on random walks. \textit{Social Networks, 27}(1), 39-54.

\bibitem{barabasi}
Barab\'{a}si, A.\ \& Albert, R.\ (1999). Emergence of Scaling in Random Networks. \textit{Science, 286}(5439), 509-512.

\end{thebibliography}

\end{document}